\documentclass[letter]{aa} 
\usepackage{graphicx}
\usepackage{natbib}
\bibpunct{(}{)}{,}{a}{}{;}
\usepackage{amssymb}
\usepackage{amsmath}
\usepackage{url}
\usepackage{units}
\usepackage{multirow,color}
\usepackage{txfonts}
\def\arcsec{$^{\prime}$$^{\prime}$}
\def\arcmin{$^{\prime}$}

\def\kms{\,km~s$^{-1}$}

\begin{document}
   \title{ALMA CO J\,=\,6$-$5 observations of IRAS16293--2422}
   \subtitle{Shocks and entrainment}


\author{L.E. Kristensen\thanks{\emph{Present address:} Harvard-Smithsonian Center for Astrophysics, 60 Garden Street, MS78, Cambridge, MA 02138, USA, \email{lkristensen@cfa.harvard.edu}}
\and P.D. Klaassen
\and J.C. Mottram
\and M. Schmalzl
\and M.R. Hogerheijde
}

\institute{
Leiden Observatory, Leiden University, PO Box 9513, 2300 RA Leiden, The Netherlands
}

\date{Submitted: Oct. 31, 2012; Accepted: Dec. 7, 2012}


\abstract
{Observations of higher-excited transitions of abundant molecules such as CO are important for determining where energy in the form of shocks is fed back into the parental envelope of forming stars. The nearby prototypical and protobinary low-mass hot core, IRAS16293--2422 (I16293) is ideal for such a study. The source was targeted with ALMA for science verification purposes in band 9, which includes CO J=6--5 ($E_{\rm up}$/$k_{\rm B}$$\sim$116 K), at an unprecedented spatial resolution ($\sim$0\farcs2, 25 AU). I16293 itself is composed of two sources, A and B, with a projected distance of 5\arcsec. CO J=6--5 emission is detected throughout the region, particularly in small, arcsecond-sized hotspots, where the outflow interacts with the envelope. The observations only recover a fraction of the emission in the line wings when compared to data from single-dish telescopes, with a higher fraction of emission recovered at higher velocities. The very high angular resolution of these new data reveal that a bow shock from source A coincides, in the plane of the sky, with the position of source B. Source B, on the other hand, does not show current outflow activity. In this region, outflow entrainment takes place over large spatial scales, $\gtrsim$100~AU, and in small discrete knots. This unique dataset shows that the combination of a high-temperature tracer (e.g., CO J=6$-$5) and very high angular resolution observations is crucial for interpreting the structure of the warm inner environment of low-mass protostars.}


\maketitle

\section{Introduction}

\begin{figure*}
\begin{minipage}{8.5cm}
\center\includegraphics[width=7.7cm]{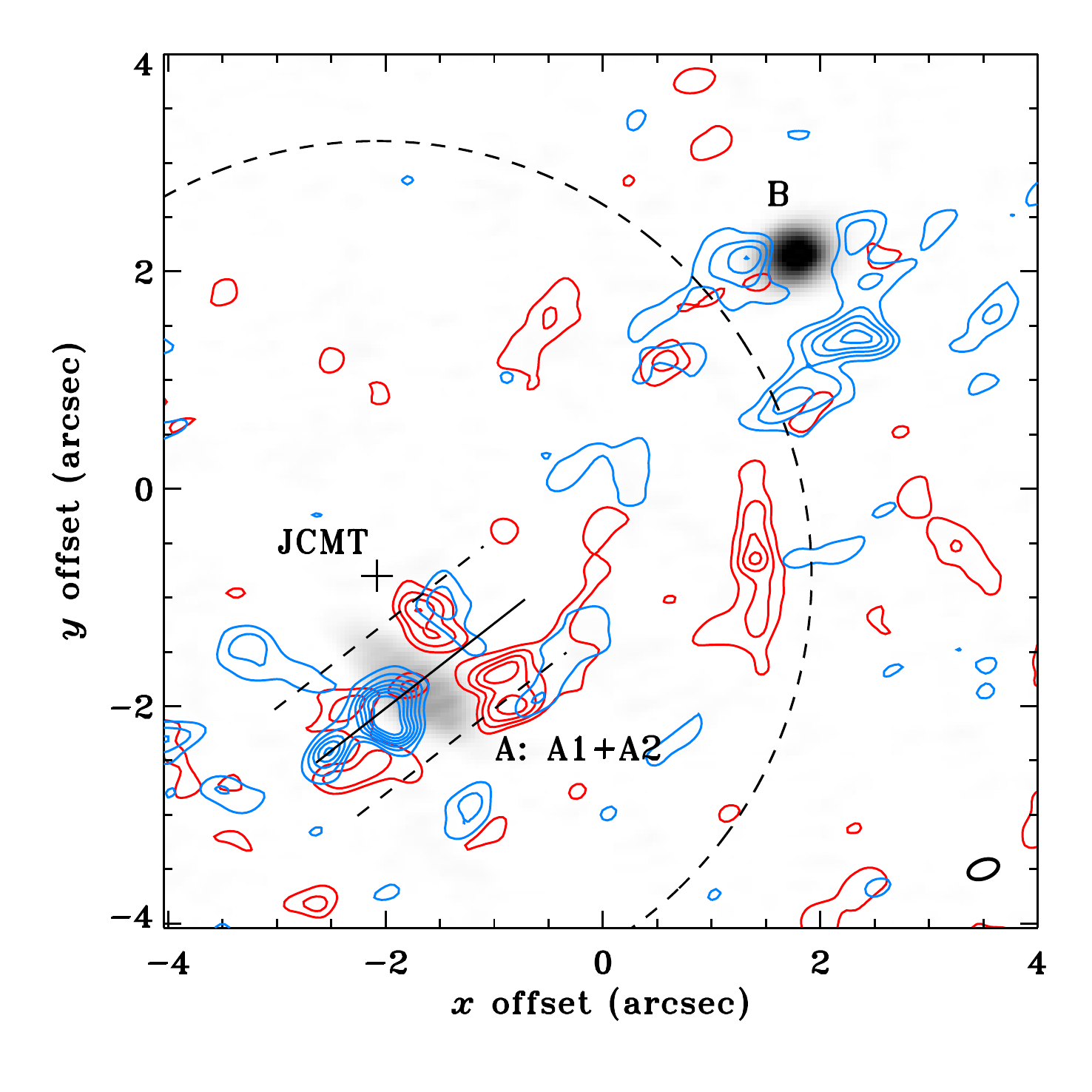}
\end{minipage}
\begin{minipage}{8.5cm}
\center\includegraphics[width=7.7cm]{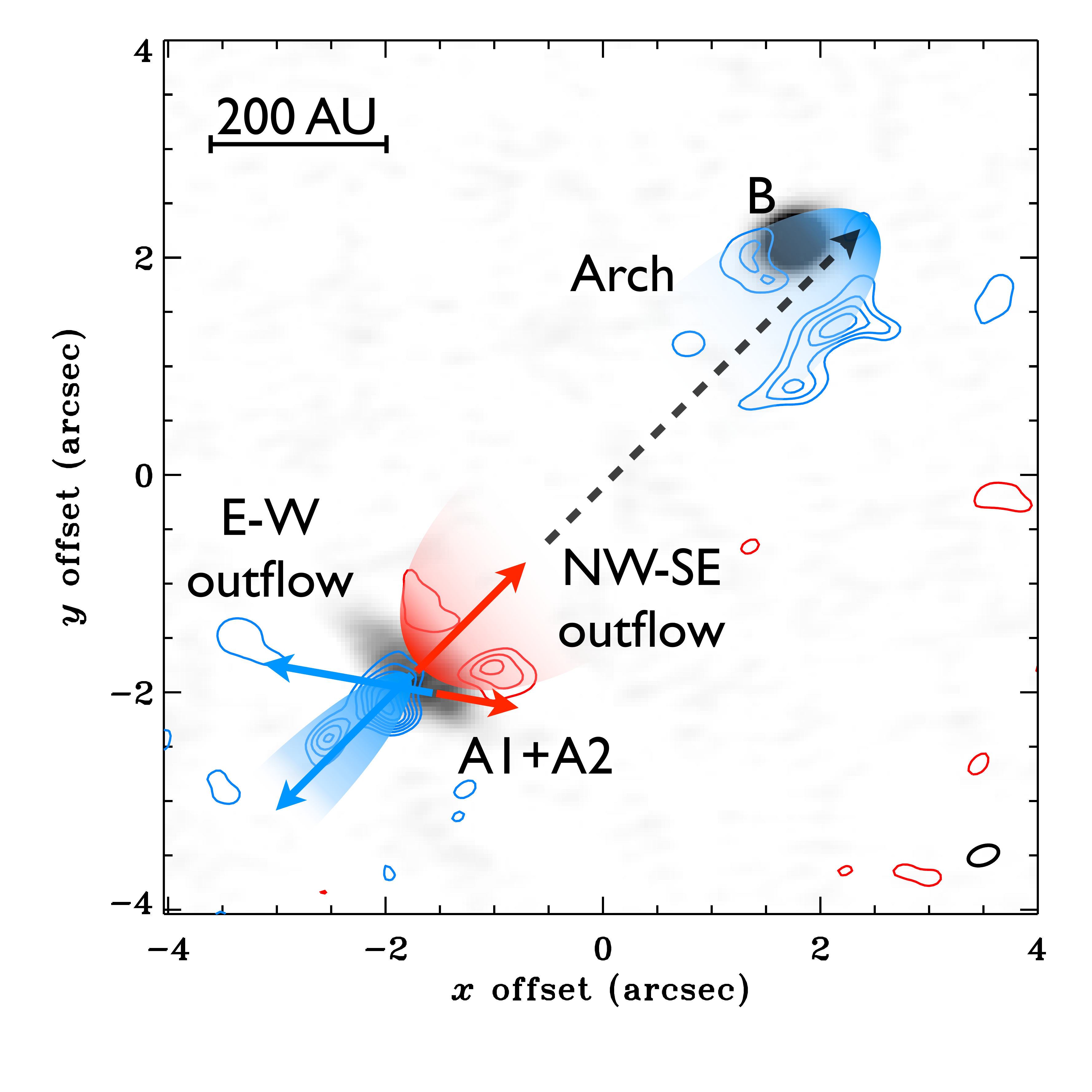}
\end{minipage}
\caption{\emph{Left:} Map of CO J=6$-$5 emission towards I16293 with contours at 6, 12, 18, \ldots $\sigma$ of integrated intensity. Integration limits are $-$10 to $+$1\kms{} and $+$7 to $+$18\kms{} for the blue- and red-shifted emission, respectively. The underlying gray-scale image shows continuum emission at 690 GHz. The positions of sources A and B are marked; source A consists of the two sources A1 and A2. Offsets are recorded from 16$^{\rm h}$32$^{\rm m}$22\fs753; --24\degr28\arcmin34\farcs747 (J2000). The ALMA beam is shown in the lower right corner (black). The centre of the JCMT beam is marked with a cross and the beam is shown as a dashed circle. The dashed lines on source A shows the region used to construct the pv diagram (Fig. \ref{fig:pv}). \emph{Right:} The emission integrated over the highest velocities (from $-$10 to $-$4\kms{} and $+$12 to $+$18\kms) is shown in contours at 5, 10, 15, \ldots $\sigma$. The different features are highlighted by arrows and are labeled. The black dashed arrow is an extrapolation of the red lobe of the NW-SE outflow.}
\label{fig:map}
\end{figure*}

IRAS16293$-$2422 (hereafter I16293) is a well-studied solar-type protostar, originally singled out as having one of the most prominent outflows of all the IRAS sources associated with the $\rho$ Ophiuchus cloud, and showing indications of infall \citep{Walker1986, Walker1988}. Sub-millimetre and radio observations reveal that the source breaks up into two sources, A and B, located at a projected distance of 5\arcsec{} \citep{Mundy1986, Wootten1989}. Source A itself breaks up into two sources, A1 and A2 separated by 0\farcs3 \citep{Wootten1989}. At the distance to the source \citep[120\,pc,][]{loinard08}, these angular separations correspond to physical distances of $\sim$600\,AU and 36\,AU respectively. While sources A1 and A2 appear active and drive two outflows, source B is more quiescent. Infall is clearly detected towards source B \citep[e.g.,][]{Pineda2012} and several shock tracers such as SiO are detected \citep[e.g.,][]{jorgensen11}, although other observations indicate source B is a prestellar core \citep{chandler05} or a T Tauri star \citep{stark04, takakuwa07}. Figure \ref{fig:map} shows the locations of the different protostellar components.

Of the two sources A and B, source B has been the more puzzling because molecular lines observed towards the source show a large spread in $\Delta\varv$ ranging from $\sim$ 1 to 5 \kms{} \citep{jorgensen11}. Towards source A, on the other hand, the lines all have a narrow range of $\Delta\varv$ of $\sim$ 2$-$3\kms{}. A high angular resolution molecular line survey carried out with the Submillimeter Array (SMA) shows that several species peak towards both sources whereas some are only detected towards source A, others only towards B  \citep{bisschop08, jorgensen11}. ALMA observations towards source B have recently shown infall in CH$_{3}$OCHO \citep{Pineda2012} and the first detection of the simplest sugar, glycolaldehyde, as well as several related molecules \citep{Jorgensen2012}. The different morphologies and velocity structures revealed by different species make I16293 one of the most enigmatic low-mass protostellar sources as illustrated by the wealth of data available.

Understanding a protostellar system such as I16293 requires not only an understanding of the infalling envelopes, but also when and where energy is released back into the envelope, primarily in the form of shocks driven by the protostellar outflow and jet. On large scales ($>$\,10$^4$ AU) I16293 drives a quadrupolar outflow \citep{mizuno90, stark04}, but disentangling the effects of the driving jet on small scales is not straightforward. Shock hotspots created by the outflow have the potential to change the local physical conditions (density and temperature), and are best traced by species which trace currently shocked gas and are unaffected by chemistry.

In the following letter, we present ALMA science verification data of the CO J=6$-$5 transition towards I16293. The very high angular resolution of these Band 9 observations (0\farcs2 at 690 GHz), and a high-temperature tracer only weakly affected by chemistry are a perfect combination to re-examine the origin of molecular emission towards these two sources. Although CO J=6$-$5 also traces swept-up outflow gas on larger spatial scales \citep[$>$\,1000\,AU;][]{vanKempen2009, Yildiz2012}, this emission is filtered out by the interferometer and therefore these new data show only the small-scale structure where the outflow is interacting with the envelope.

\section{Observations and results}
\label{sec:obs}

The observations consist of a seven-point mosaic in Band 9, and were taken as one of the ALMA Science Verification datasets on April 16 (1 execution) and 17, 2012 (3 executions) using 15 antennas and a configuration with baselines ranging from 26 to $\unit[402]{m}$ (62 to $\unit[943]{k\lambda}$ at $\unit[690]{GHz}$). The on-source integration time was $\sim$18~minutes per pointing and the primary beam is 9\arcsec. All four spectral windows consist of 3840 channels with a channel separation of $\unit[488]{kHz}$ ($\unit[0.21]{km\,s^{-1}}$ at 690 GHz) and a total bandwidth of 1.875 GHz. The calibrated data were taken from the \textsc{CASA} guides website\footnote{\url{http://casaguides.nrao.edu/}}. We performed phase and amplitude self calibration on the continuum, confirming the results from the \textsc{CASA} guide. The gain solution from the self calibration was then applied to the measurement set, and the spectral region around the CO 6--5 line at 691.473 GHz was extracted. 

Due to the combination of large-scale emission of CO and the missing short spacings, special care was taken to identify and flag data points with amplitude outliers. These could result in image artifacts being interpreted as real structures due to incomplete sampling of the $uv$-plane. Since the $uv$-coverage of the dataset obtained on April 16 has a large overlap with one of the datasets from April 17, the derived amplitudes were compared to help identify bad data. Two out of 105  baselines exhibited questionable amplitudes and were therefore flagged. The details of this procedure are discussed in Appendix \ref{app:dr}. The visibilities were imaged using a robust weighting of 0.5. The resulting spatial resolution is 0\farcs29$\times$0\farcs16 at a position angle of $\unit[110]{\degr}$. The rms noise is $\unit[100]{mJy\,beam^{-1}}$ per 0.21\kms{} channel. The spectral cube and the continuum image were corrected for primary-beam attenuation, and smoothed to an effective resolution of $\unit[0.84]{km\,s^{-1}}$.

\begin{figure}
\center\includegraphics[width=0.9\columnwidth]{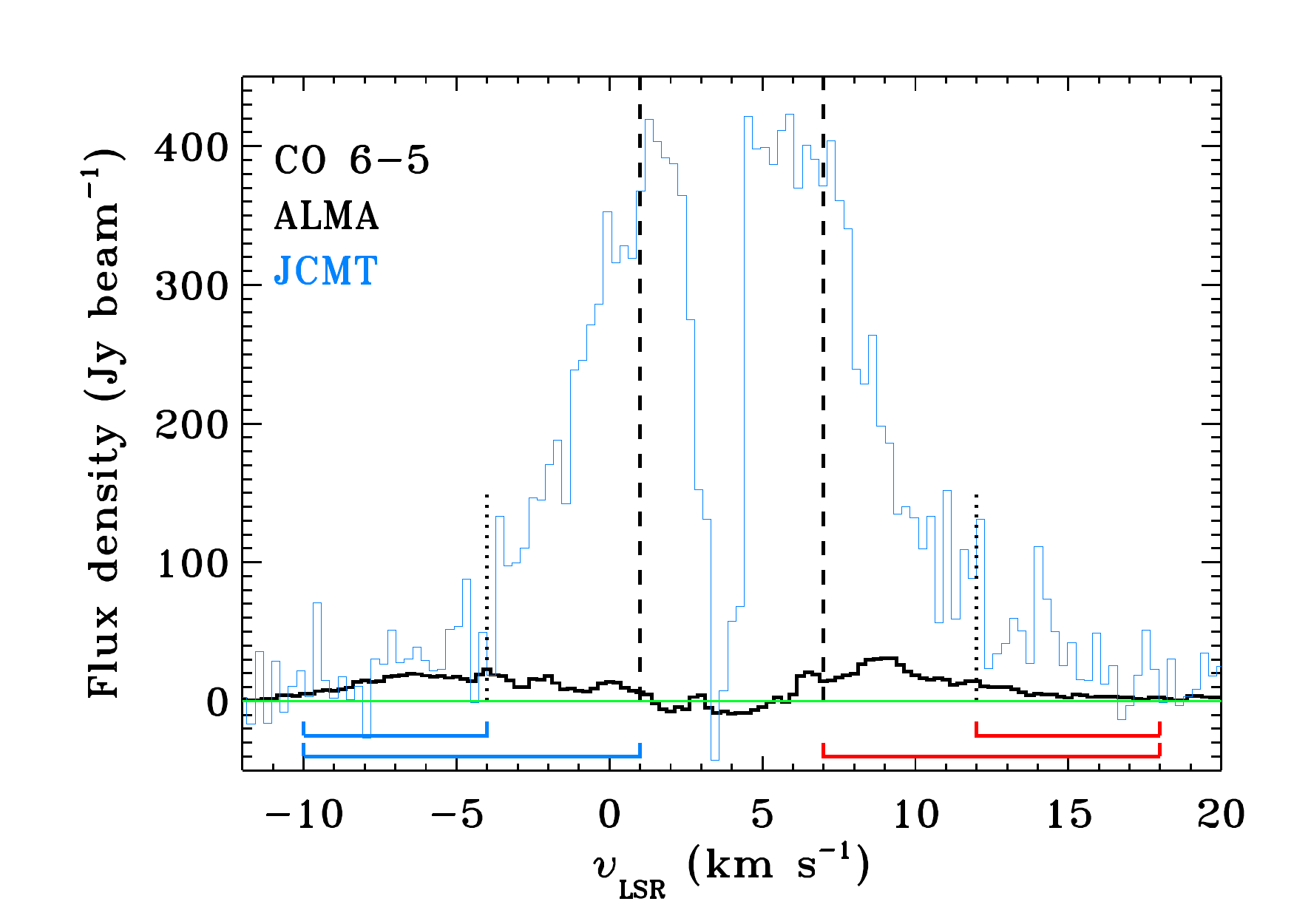}
\caption{ALMA (black) and JCMT (blue) continuum-subtracted CO J=6--5 spectra compared in the same 8\arcsec{} beam marked in Fig. \ref{fig:map}. The dashed lines indicate the velocity interval excluded in this analysis ($\varv_{\rm source}$\,$\pm$3\,km\,s$^{-1}$) and the dotted lines illustrate the high-velocity intervals less affected by filtering. The horizontal line (green) is the baseline.}
\label{fig:filtering}
\end{figure}

CO J=6$-$5 emission is clearly detected and the integrated emission shows complex structure. Figure~\ref{fig:map} shows CO J=6$-$5 emission overlaid as contours on the continuum, where the blue and red-shifted emission is integrated separately from $-$10 to $+$1\kms{} and $+$7 to 18\kms{} respectively (Fig. \ref{fig:filtering}). As can be seen from Fig. \ref{fig:map}, there is very little overlap between blue- and red-shifted emission except around source A where there is some overlap. 

To quantify the amount of spatial filtering in these data, we compare emission to a CO J=6--5 spectrum from the James Clerk Maxwell Telescope (JCMT) towards I16293-A. The spectrum was obtained under good atmospheric conditions in July 2000 and is similar to that presented in \citet{stark04}. The JCMT was pointed to a position approximately 1\arcsec{} N of source A and the beam-size is 8\arcsec{} (Fig. \ref{fig:map}). The JCMT data were converted to flux density using a conversion factor of 34.0 Jy\,beam$^{-1}$ K$^{-1}$. The ALMA data were convolved with an 8\arcsec\ beam centred at the same position (Fig. \ref{fig:filtering}). ALMA recovers $\sim$5\% of the flux density in the full velocity range, but when considering only the blue or red wings (excluding the central 6\kms{} from $\varv_{\rm LSR}$ $\pm$ 3\kms), approximately 15 and 10\% of emission is recovered, respectively. If the velocity cut-off  is moved from $\varv_{\rm source}$\,$\pm$\,3\kms{} to $\pm$\,8\kms{}, $\sim$\,50 and 15\% of emission is recovered in the blue and red wings, respectively. These high-velocity structures are therefore less affected by filtering. They are shown in Fig. \ref{fig:map} (right) where they likely form the edges of larger-scale structures not recovered.

\begin{figure}
\center\includegraphics[width=0.87\columnwidth]{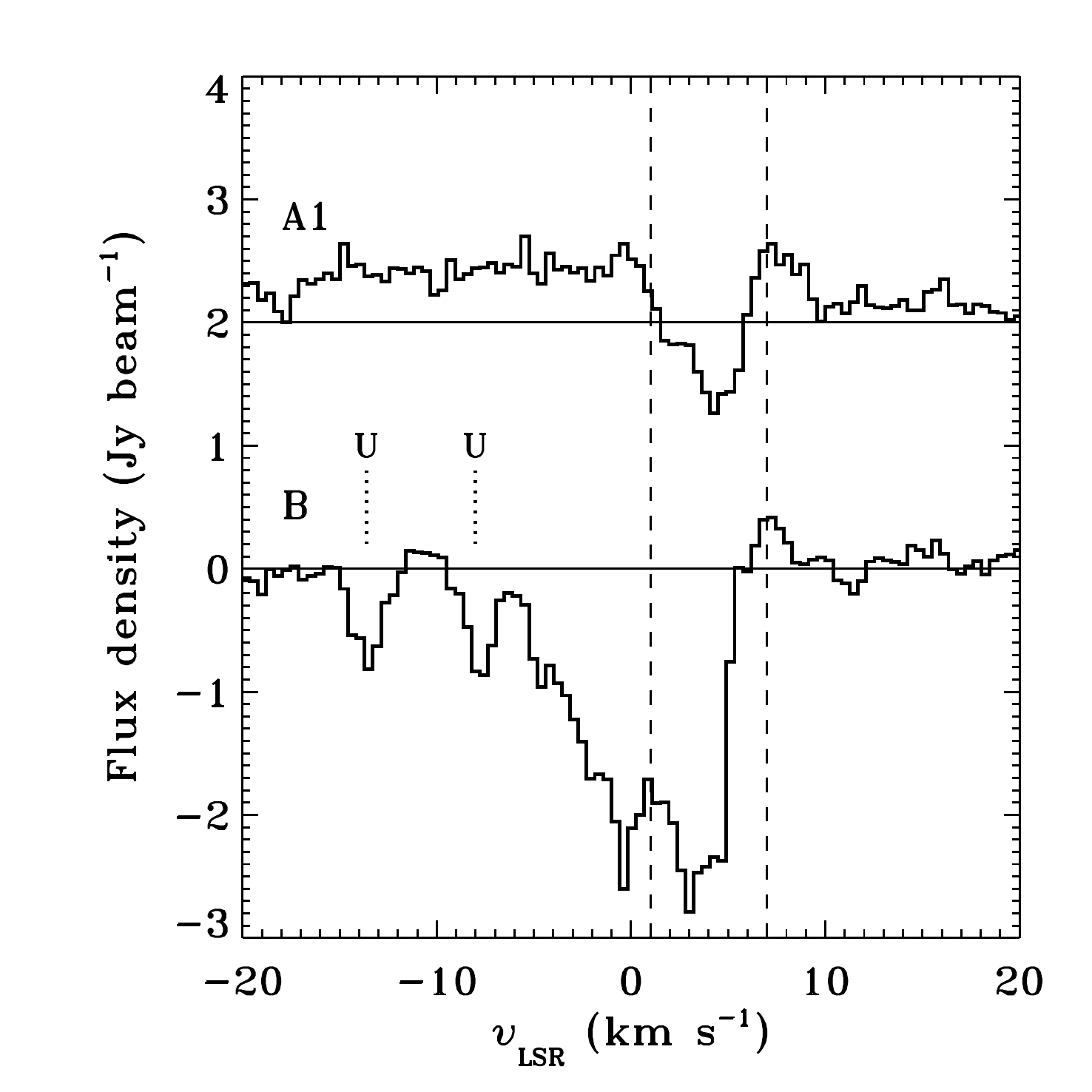}
\caption{Continuum-subtracted spectra towards the sources A and B. The spectrum towards A has been shifted by 2 Jy beam$^{-1}$. Baselines are shown as full lines, and the velocity range excluded from analysis is shown with vertical dashed lines. Unidentified features are marked with a U (see text).}
\label{fig:spectra}
\end{figure}

The peak flux densities are typically of the order 1$-$2 Jy per ALMA beam. Figure \ref{fig:spectra} shows single-point spectra towards the continuum peaks of sources A and B. Towards source A, the profile is dominated by blue-shifted emission and absorption at the source velocity. Towards source B, the CO J=6$-$5 profile is seen almost entirely in absorption against the strong continuum. More features are seen in absorption towards source B, similar to the observations of glycolaldehyde \citep{Jorgensen2012}; we do not speculate on the nature of the species responsible for the two absorption features labeled ``U'' in Fig. \ref{fig:spectra}.

\section{Discussion: origin of the emission}
\label{sec:disc}

The highest-velocity emission is shown in the right panel of Fig.~\ref{fig:map} where it outlines active hotspots where the outflow is interacting with the envelope. We therefore focus the discussion on their morphology and the interpretation. Source A hosts a red-shifted arch capped by blue-shifted emission pointing NW and blue-shifted emission surrounded by red-shifted emission stretching SE. The wing emission is strongest and most extended in terms of velocity towards the region SE of source A. Near source B an arch of blue-shifted emission opens to the SE towards source A. 

Concerning the knots around source A, the origin of the large-scale E-W outflow is seen in two outflow knots that are aligned with source A2. This is consistent with the observations of \citet{pech10} and \citet{loinard12}. On these small scales the outflow direction is switched in the sense that the red-shifted lobe is moving west, while the blue-shifted is moving east, opposite to what is observed on large scales \citep[e.g.,][]{stark04}. The lobes of the large-scale E-W flow do not extend continuously to the inner region, implying that the system is entering a new accretion phase with a slightly different inclination angle \citep{pech10, loinard12}. 

The distance traversed by the red-shifted shock of the E-W flow (the tip of the eastern red arrow in Fig. \ref{fig:map}), $\sim$0\farcs5, since its launch in 2005 is consistent with proper motions seen in the cm observations of this outflow knot \citep[$\sim$0\farcs1 yr$^{-1}$;][]{pech10} and implies a space velocity of $\sim$90\kms{} with an inclination angle of $\sim$5\degr\ with respect to the plane of the sky. With ALMA it is possible to trace the motion of these outflow knots on timescales of a few years. Thus, the velocity and momentum transfer can be followed in 3D, providing tight constraints on entrainment simulations \citep{arce07}.

There is a clear asymmetry between the blue- and red-shifted emission of the NW-SE outflow in terms of spatial distribution. The red-shifted emission NW of source A breaks into 3$-$4 knots that are aligned in a semi-shell. These knots may represent small shock condensations where the outflow is directly interacting with the envelope, as is also prominently traced by thermal H$_2$O emission towards other sources \citep{kristensen12} and H$_2$O maser emission towards this source \citep{wootten99}. The blue-shifted emission SE of source A however, shows a highly collimated structure bracketed by red-shifted emission.

\begin{figure}
\center\includegraphics[width=0.9\columnwidth]{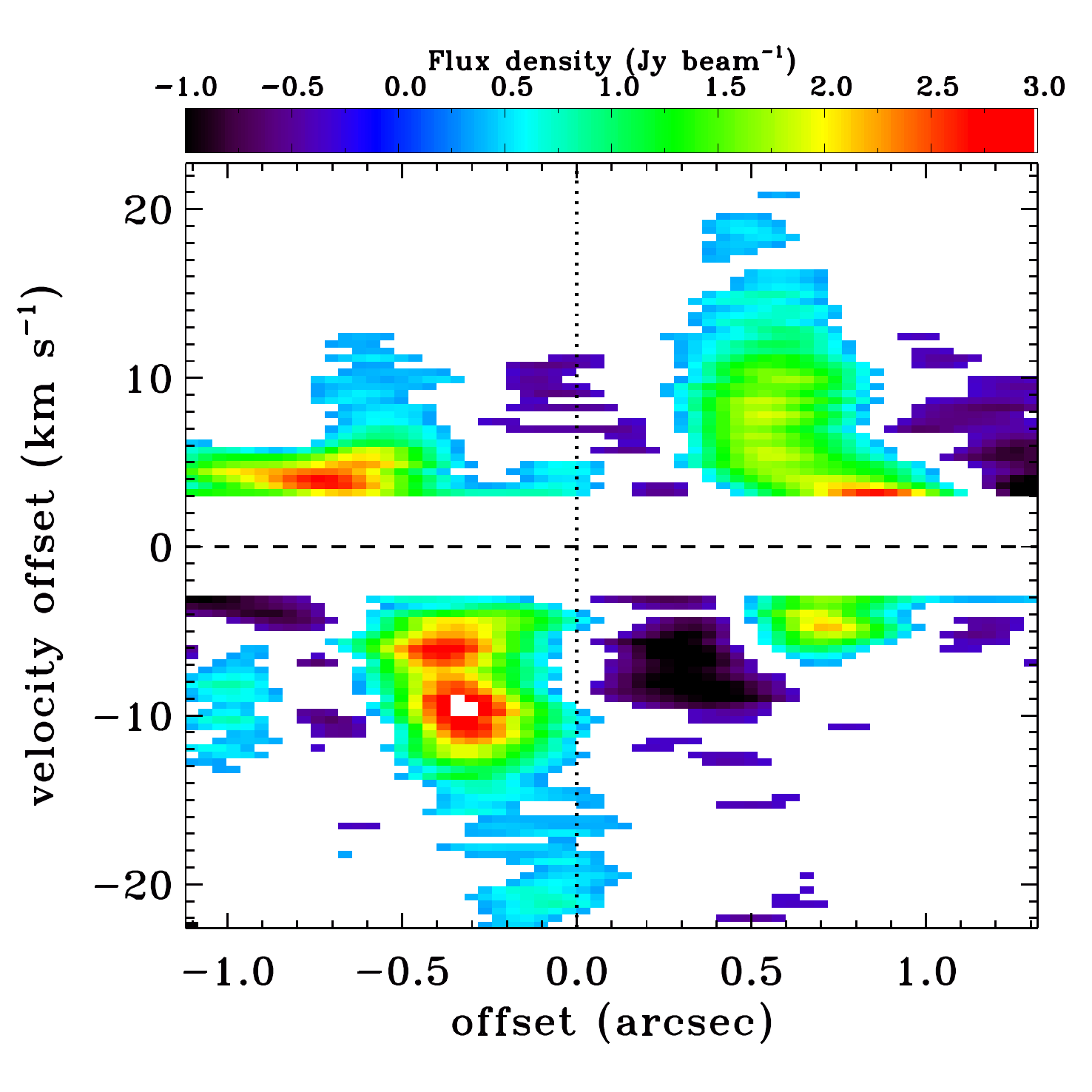}
\caption{Position-velocity diagram of a slice running at a position angle of 128\degr\ and with a width of 1\farcs25 covering the central part of the outflow centred on source A1. The velocity has been shifted so the source velocity is at 0\kms{} (black line). Signal below the 3$\sigma$ level is ignored as are the central $\pm$3 \kms{}. The position of source A1 is marked with a dotted line. Negative offsets are towards the SE, positive towards NW.}
\label{fig:pv}
\end{figure}

To examine whether the cause of the asymmetry is internal (e.g., asymmetric launch) or external (e.g., density gradient or clumps) we created a position-velocity (pv) diagram with a wide slit of 1\farcs25 to capture the red-shifted emission NW of source A (Fig. \ref{fig:pv}). The line profiles extend out to $\pm$20\kms{} in these knots. The velocity of the blue-shifted emission is highest at the base of the outflow ($\varv$ $\sim$ $-$20 \kms{}), closest to the source but rapidly decelerates over 0\farcs5 ($\sim$ 60 AU) to just a few \kms{}. Part of the red-shifted emission appears to be hidden by the continuum and is only in emission at a distance of $\sim$ 0\farcs3 (35 AU). The maximum velocity is also 20 \kms. At a distance of 1$''$ (120 AU) from the source the velocity of the red-shifted emission has also dropped to the ambient velocity. The symmetry in the pv diagram suggests that the origin of emission is the same and the cause for the asymmetry is in the surrounding environment, but on small scales ($\lesssim$ 25 AU). 

The rapid deceleration seen in the pv diagram is different to the Hubble-type flows observed over larger spatial scales \citep[][and references therein]{arce07} but more consistent with a ``turbulent wind''-type scenario \citep{canto91} where the outflow material decelerates through interaction with the surrounding envelope. Furthermore, the pv diagram illustrates that entrainment takes place over scales larger than the inner $\sim$ 100 AU, at least in this system, and in localised hotspots.

The arch near source B delineates an empty half-shell; the arch likely contains large-scale filtered emission---full ALMA  (with the Atacama Compact Array, ACA, and Total Power, TP, telescopes included) will be required to capture this emission. This arch is a coherent structure with a characteristic line width  of $\sim$ 5\kms{} which is more consistent with the line widths of source A than B.  The arch is seen in absorption against source B (Fig. \ref{fig:spectra}), which is offset from the apex of the arch. The offset is particularly clear in individual channel maps (Fig. \ref{fig:channel_map} in Appendix \ref{app:rt}). The arch is aligned with the NW-SE outflow originating in source A1 (Fig. \ref{fig:map}). If source B were the driving source of this structure we would expect the highest-velocity gas to be closest to source B, which is not the case. If source B were the origin of the structure and the blue arch were an outflow cone, the outflow structure at the base of the outflow within 25 AU would be significantly different from that of source A (see Appendix \ref{app:rt} for a derivation). Contrary to the very recent interpretation by \citet{loinard12}, these facts are consistent with an origin in source A, and that the arch is a bow shock originating from jet interacting with the ambient material surrounding source B.

At the base of the outflow near source A itself, both red- and blue-shifted emission is observed and overlapping as can be seen both in the map (Fig. \ref{fig:map}) and the pv diagram (Fig. \ref{fig:pv}), indicating that the outflow is moving close to the plane of the sky. \citet{chandler05} reached a similar conclusion based on the large-scale outflow maps of \citet{stark04}. Even small-scale precession in and out of the plane of the sky could cause the arch to be blue-shifted at a distance of $\sim$ 5\arcsec{} (600 AU). The dynamic age of the blue arch is $\lesssim$ 300 yrs for a flow velocity of 10\kms{} (Fig. \ref{fig:spectra}), an upper limit on the age because the flow velocity is a lower limit. Therefore, any precession in and out of the plane will have to be on timescales shorter than $\sim$ 300 yrs. \citet{pech10} demonstrate that A1 and A2 are a binary system and \citet{chandler05} find that the position angle between the two sources varies by 2.2\degr{} yr$^{-1}$ whereas the projected distance remains constant, implying an orbital period of $\sim$ 165 yrs. It is therefore plausible that the binary interaction is responsible for some of the difference in motion with respect to the plane of the sky.

Several species have been detected inside the arch with the Submillimeter Array (SMA) \citep{chandler05, takakuwa07, jorgensen11}, including SiO and SO which are both known shock tracers. The SiO channel maps by \citet{jorgensen11} show that the SiO emission is both blue- and red-shifted, and SMA maps of CO J=2$-$1 and J=3$-$2 show that there is significant blue-shifted emission towards source B \citep{takakuwa07, yeh08}. These observations corroborate the scenario that the arch is an active shock region. This interpretation is also consistent with the large range of $\Delta\varv$ observed towards source B: some of the species are not associated with this source but arise in the shock originating from source A along the same sightline. In particular, the lines with a width of $\sim$1\kms{} likely originate in the inner protostellar envelope, whereas the broader lines ($\sim$5\kms) originate in this bow shock. Source B itself appears to be quiescent at the moment and is not actively accreting although larger-scale infall is observed towards the source \citep{Pineda2012}.

\section{Summary and conclusions}

CO J=6--5 is detected in a large number of distinct clumps towards I16293 where shocks are likely active. More flux density is recovered in the high-velocity line wings by the interferometer, and so it is not surprising that only localised higher-velocity hotspots are detected. The new data show that there is no direct evidence for current shock activity associated with source B,  and that although infall is observed towards the source, it is not currently accreting. Instead, a blue-shifted bow shock from source A is overlapping with source B in the plane of the sky. Outflow entrainment takes place over large scales, $\gtrsim$100 AU, and wind material is decelerated through direct interaction with the envelope. These ALMA science verification data already reveal the potential of this new facility for shedding light on how protostars interact with the dense inner envelope on spatial scales of $\lesssim$100 AU and where energy is fed back into the envelope. 

\begin{acknowledgements}
This paper makes use of the following ALMA data: ADS/JAO.ALMA\#2011.0.00007.SV. ALMA is a partnership of ESO (representing its member states), NSF (USA) and NINS (Japan), together with NRC (Canada) and NSC and ASIAA (Taiwan), in cooperation with the Republic of Chile. The Joint ALMA Observatory is operated by ESO, AUI/NRAO and NAOJ. Astrochemistry is Leiden is supported by NOVA, a Spinoza grant, grant 614.001.008 from NWO, and by EU FP7 grant 238258. Allegro, the ALMA Regional Centre node in the Netherlands, is supported by NOVA and NWO. We are grateful to T. van Kempen and the JAO CSV for planning and taking these data, and to E.F. van Dishoeck and S. Cabrit for very fruitful discussions. We would like to thank the referee, G. Fuller, for constructive comments which lead to  improvement of this paper.
\end{acknowledgements}

\bibliography{bibliography}

\appendix

\section{Data reduction details}\label{app:dr}

\begin{figure*}
\sidecaption
\includegraphics[width=12cm]{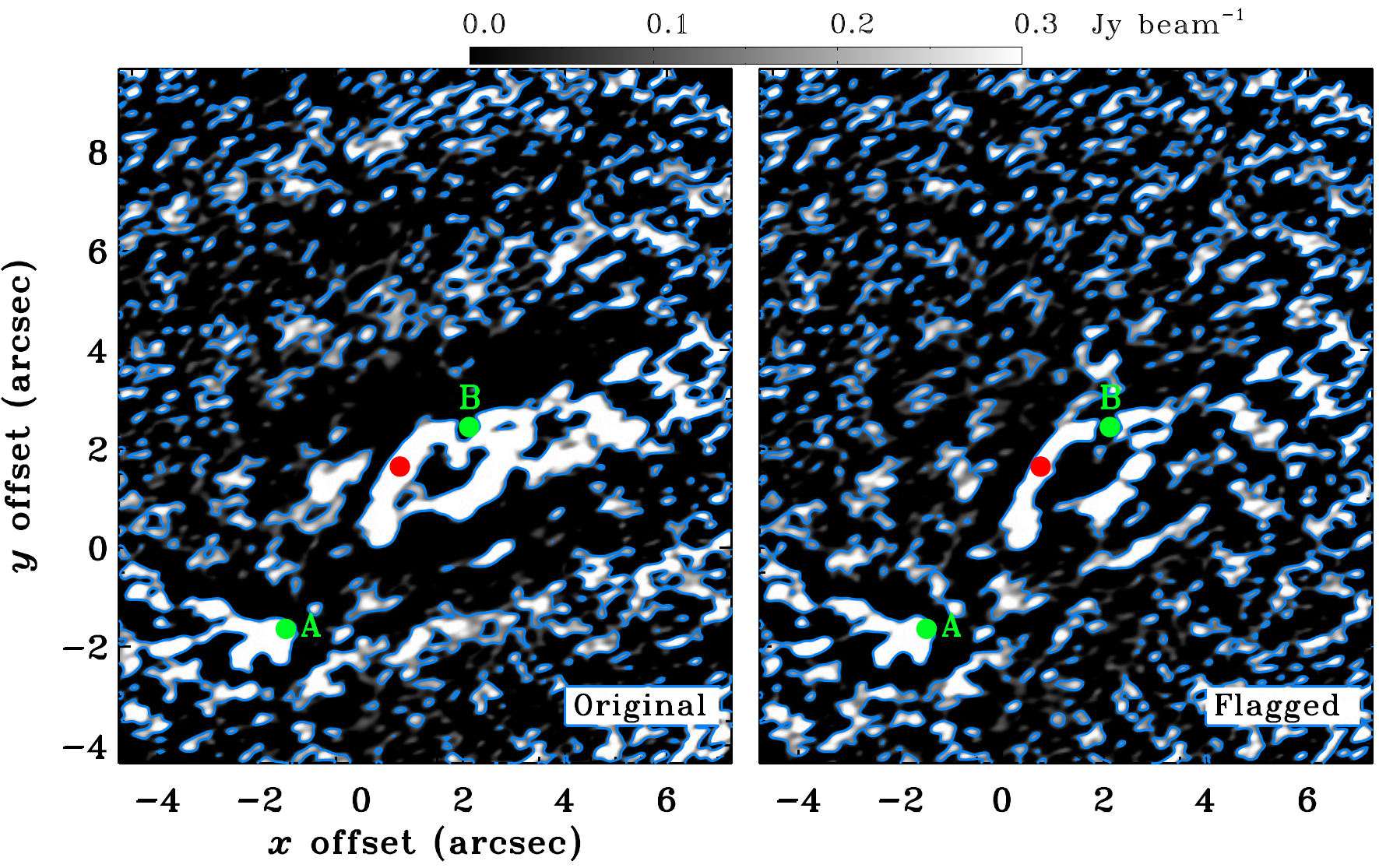}
\caption{Channel maps at $-$3\kms{} before and after additional flagging (see text). The offsets are as in Fig. \ref{fig:map}. The color scale ranges from 0 to 0.3 Jy beam$^{-1}$ (0 to $\sim$ 3$\sigma$) and has been chosen to highlight the striping present in the left map (original data) but not the right (flagged data). The blue contours are at 1$\sigma$ in this 0.25\kms{} channel. The red dot at (0\farcs75, 1\farcs65) shows the position where the spectrum in Fig. \ref{fig:specm} was extracted. Sources A and B are marked in green.}
\label{fig:pre_post}
\end{figure*}

For extended emission---as seen in CO J=6--5 towards IRAS16293---the $uv$ sampling density towards short baselines is crucial for a successful recovery of large spatial scales. In the image reconstruction process, each baseline contributes a corrugation pattern along a position angle in the image plane that corresponds to the position angle of the antenna pair in the $uv$ plane. Amplitude and position of this wave-like pattern are determined by its complex visibility. A single baseline, therefore, does not provide enough information to constrain the spatial origin of the emission. The combined information of many antenna pairs with various baseline lengths and position angles is required to get an image with genuine structures.

This also means that if a certain region in the $uv$ plane is under-sampled, this will lead to imaging artifacts. To a certain degree these can be removed during the image deconvolution process. However, if an under-sampled baseline exhibits strong correlated flux density, the corrugation pattern, which is caused by this antenna pair, will remain in the deconvolved image and might be misinterpreted as real structure. For this reason we flag the baseline DA47\&DV14, which is the only one that samples spatial scales above 3\arcsec. Since this baseline exhibits high correlated flux densities, the reconstructed image shows a clear corrugation pattern, which is indicative of extended structures that cannot be quantified correctly (Fig. \ref{fig:pre_post}).

Additional analysis of the visibilities lead to the flagging of the antenna pair DA41\&DV09. We made use of the fact that two execution blocks were observed almost exactly 24 hours apart, leading to almost identical $uv$ coverage. This made it possible to cross check the visibilities. In certain channels at blue-shifted velocities, DA41\&DV09 exhibited strong correlated flux densities, but with significant differences between the two execution blocks. 

\begin{figure}
\center\includegraphics[width=\columnwidth]{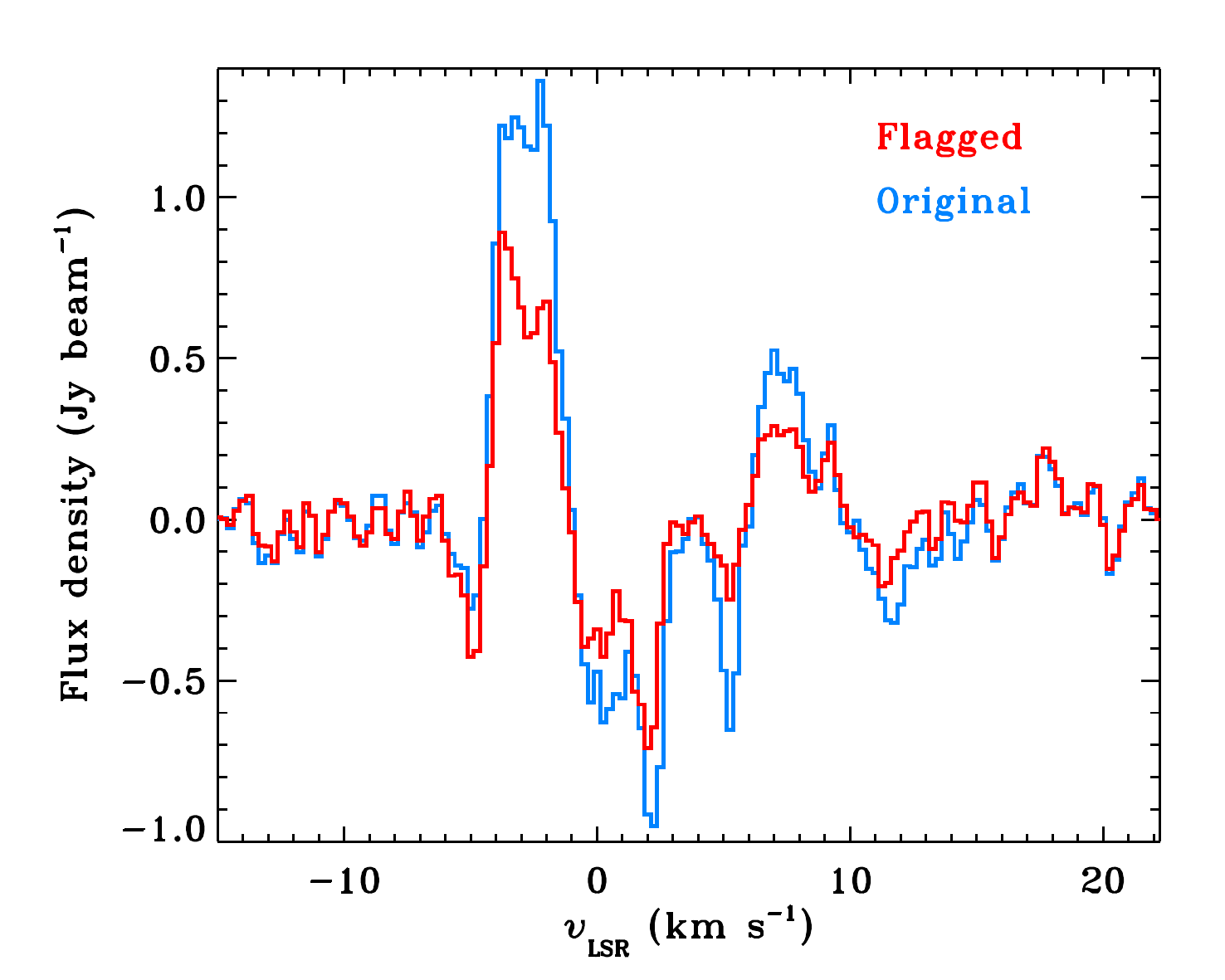}
\caption{Spectrum towards a position located along the blue arch and marked in Fig. \ref{fig:pre_post}. The flagged and original data are shown in red and blue, respectively.}
\label{fig:specm}
\end{figure}

The flagging of these two baselines significantly changed the spectra at certain positions in the sky (see Fig. \ref{fig:specm} for an example), particularly at positions associated with the blue arch. We suspect that large-scale structure ($>$3\arcsec) is associated with the blue arch, but that full ALMA is required to recover it (with the Atacama Compact Array, ACA, and Total Power, TP, telescopes included). We note that the overall morphology of the system has not changed, and all key structures can be recovered although the blue arch appears to sit on a plateau of emission at velocities of $\sim$3\kms{} in the original data (Fig. \ref{fig:pre_post}).

\section{Comparison of outflow emission at sources A and B}\label{app:rt}

We illustrate below that if the arch were caused by source B, instead of being a chance alignment, a very different outflow \emph{collimating} structure would be required to that found in source A. We do so by assuming that the opacity towards the continuum peaks of sources A and B are the same ($\tau_v^A$ = $\tau_v^B$ = $\tau$) and that the line emission source functions are the same ($S_v^A$ = $S_v^B$ = $S_v$) towards both continuum peaks, but that only a fraction, $f$, of the emission will be recovered by the interferometer. This fraction may be different towards the two sources, and we denote the fraction recovered towards source A with $f$ and towards source B with $\eta$$f$. Only in the case where the outflow structure at the base of the outflow (i.e., at the continuum peak) is similar towards the two sources is $\eta$ = 1.

The standard equation of radiative transfer reads:
\begin{equation}
I_v = I_v(0)\,e^{-\tau_v} + S_v\, (1-e^{-\tau_v}) \,
\end{equation}
where $I_v$(0) is the continuum contribution. Towards sources A and B, respectively, the radiative-transfer equations become:
\begin{align}
{\rm A:}\ I_v^{\rm A} &= I_v^{\rm A}(0)\,e^{-\tau_v} + f\,S_v\, (1-e^{-\tau_v}) \\
{\rm B:}\ I_v^{\rm B} &= I_v^{\rm B}(0)\,e^{-\tau_v} + \eta\,f\,S_v\, (1-e^{-\tau_v})\ .
\end{align}
When moving into the blue-shifted line wing towards source A, the line emission decreases until almost nothing is left but the continuum:
\begin{equation}
I_v^{\rm A} = I_v^{\rm A}(0)\,e^{-\tau_v} + f\,S_v\, (1-e^{-\tau_v}) \approx I_v^{\rm A}(0)\ . \label{eq:4}
\end{equation}
In other words, line emission and absorption are balanced and it follows from Eq. \ref{eq:4} that
\begin{equation}
I_v^{\rm A}(0)\,(1-e^{-\tau_v}) \approx f\,S_v\, (1-e^{-\tau_v})\ .
\end{equation}

\begin{figure}
\center\includegraphics[width=\columnwidth]{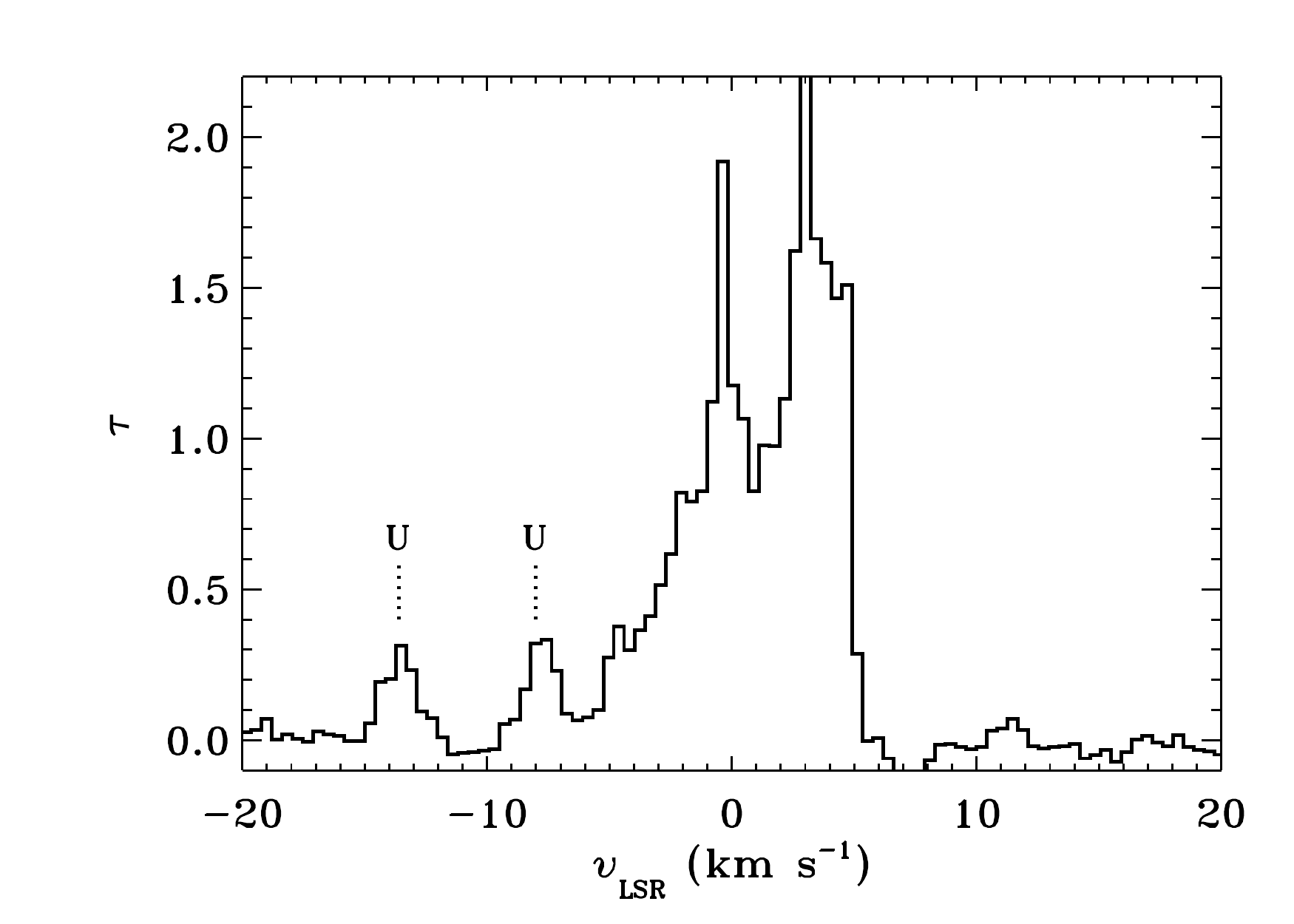}
\caption{Line opacity, $\tau$ towards the continuum peak of source B. The unidentified line features are marked with U. The opacity is a lower limit as discussed in the text.}
\label{fig:tau_b}
\end{figure}

\begin{figure}
\center\includegraphics[width=\columnwidth]{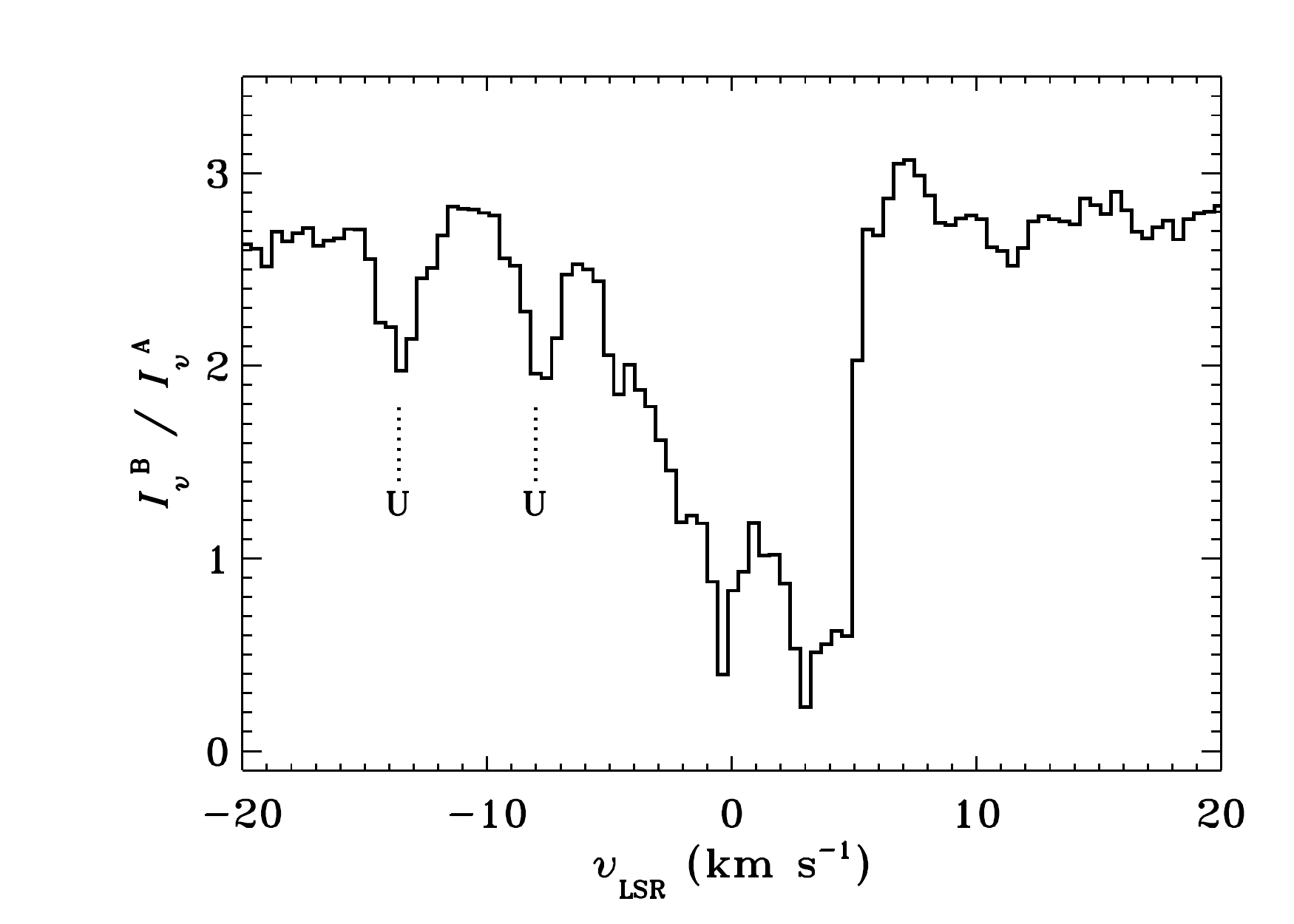}
\caption{Line emission ratio from the peak of continuum emission towards source B and a line-free position near the peak of source A. The unidentified line features are marked with U.}
\label{fig:ratio_ab}
\end{figure}

\begin{figure}
\center\includegraphics[width=\columnwidth]{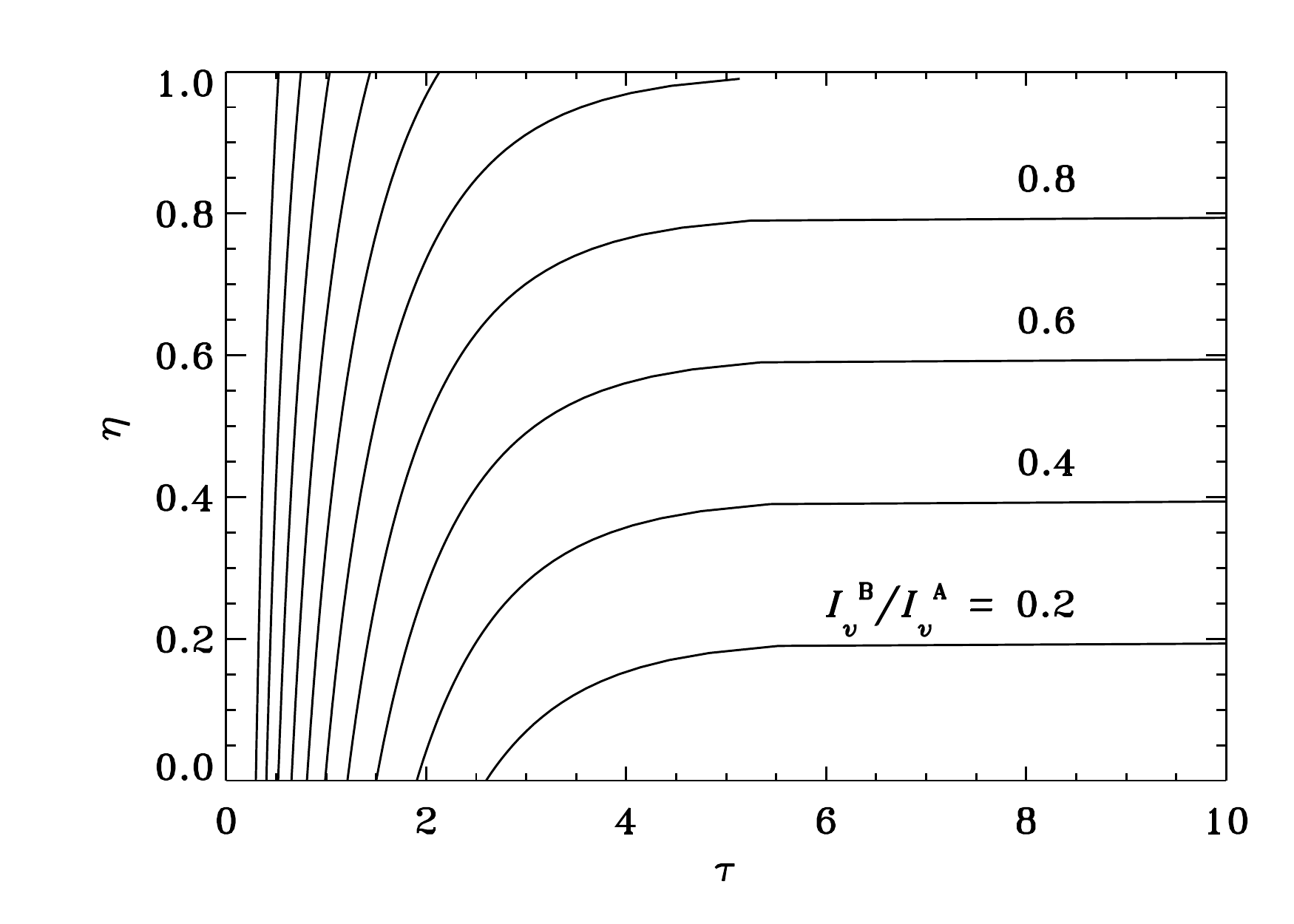}
\caption{$\eta$ as a function of the line opacity for different emission ratios. The contours are at emission ratios of 0.2, 0.4, 0.6, \ldots to 2.0 with the four lowest ratios labeled.}
\label{fig:eta_tau}
\end{figure}

The ratio of observed emission towards sources B and A is
\begin{align}
\frac{I_v^{\rm B}}{I_v^{\rm A}} &= \frac{I_v^{\rm B}(0)\,e^{-\tau_v} + \eta\,f\,S_v\, (1-e^{-\tau_v})}{I_v^{\rm A}(0)\,e^{-\tau_v} + f\,S_v\, (1-e^{-\tau_v})} \\
&= \frac{R I_v^{\rm A}(0)\,e^{-\tau_v} + \eta\, I_v^{\rm A}(0)\, (1-e^{-\tau_v})}{I_v^{\rm A}(0)} \\
&= R e^{-\tau} + \eta\, (1-e^{-\tau_v})
\end{align}
where $R = I_v^{\rm B}(0) / I_v^{\rm A}(0)$ is the ratio of the continuum emission towards the two peaks. The value of $R$ is 3.05\,Jy\,beam$^{-1}$ / 1.13\,Jy\,beam$^{-1}$ = 2.69. Solving for $\tau$ gives
\begin{equation}
\tau = \ln \left( \frac{R - \eta}{I_v^{\rm B} / I_v^{\rm A} - \eta} \right)\ .
\end{equation}
 The value of $\tau$ can be estimated towards source B based on the available data. If the absorption is caused entirely by a foreground layer, $\tau = \ln(I_v(0) / I_v)$ which is shown in Fig. \ref{fig:tau_b}. However, the arch is primarily observed in emission around source B which means that this value of $\tau$ is a lower limit because both emission and absorption needs to be accounted for. Towards the deepest absorption feature, $\tau$ is greater than 1.0--1.5. The observed intensity ratio, $I_v^{\rm B} / I_v^{\rm A}$ is displayed in Fig. \ref{fig:ratio_ab}. At the bottom of the absorption feature, the emission ratio is $\sim$ 0.5. Figure \ref{fig:eta_tau} shows $\eta$ as a function of $\tau$ for various values of the emission ratio $I_v^{\rm B} / I_v^{\rm A}$. For an observed emission ratio of 0.5, $\eta$ is at most 0.5, irrespective of $\tau$, which implies that at most $\sim$ 50\% of the emission recovered towards source A is recovered towards source B at this velocity. If more emission were recovered, the absorption feature would be filled with the emission. 
 
\begin{figure}
\center\includegraphics[width=0.75\columnwidth]{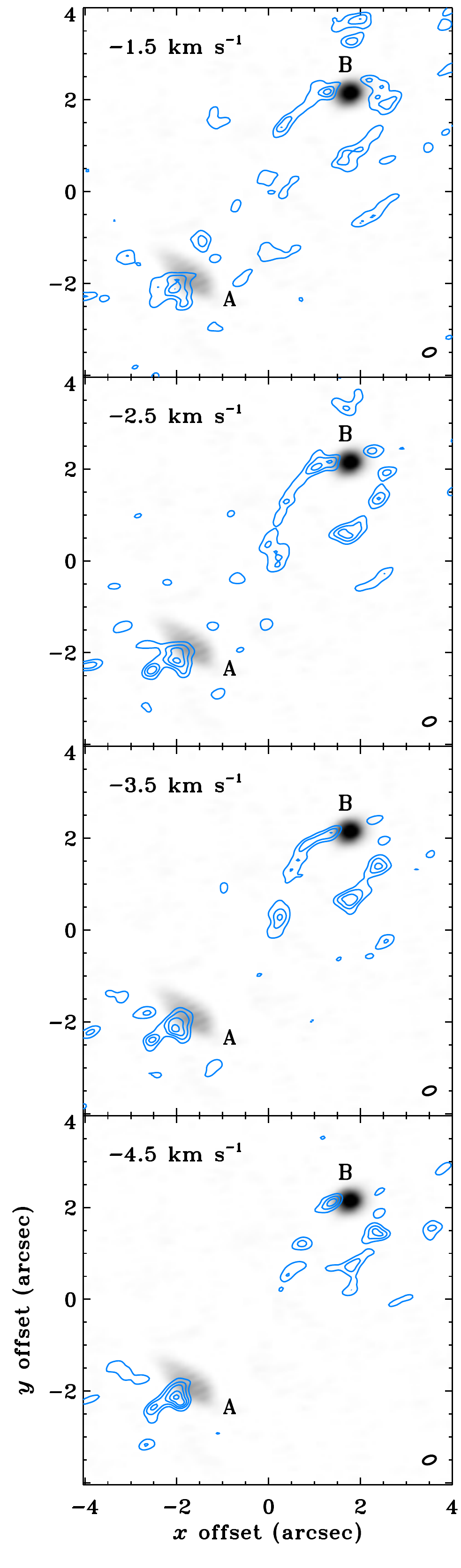}
\caption{Channel maps of blue-shifted emission at the velocities indicated in the upper left corner of each panel. The channels have a width of 0.25\kms{} and contours are at 4, 8, 12, \ldots $\sigma$. The background grayscale image is the continuum. The positions of sources A and B are marked.}
\label{fig:channel_map}
\end{figure}

The above derivation is only valid if the bases of the outflows are similar. As \citet{loinard12} argue, source B may be a candidate first hydrostatic core in which case the outflow is expected to be less collimated and therefore may be subject to more filtering. However, for that to be the case, source B must be located at the apex of the arch. Figure \ref{fig:channel_map} shows a set of CO channel maps obtained at blue-shifted velocities. All four maps, and in particular those at --1.5 and --2.5\kms, show that source B is not at the apex of the arch. From these arguments we conclude that the blue arch is unrelated to source B and is a chance alignment in the plane of the sky.

\end{document}